\documentclass[11pt]{iopart}
\usepackage{iopams}
\begin{document}

\paper[Replicon modes and stability of critical behaviour]{Replicon modes and
       stability of critical behaviour of disordered systems with respect to
       the continuous replica symmetry breaking}
\author{A A Fedorenko}
\address{Martin-Luther-Universit\"{a}t Halle-Wittenberg,
Fachbereich Physik, D-06099\\ Halle, Germany}
\ead{fedorenko@physik.uni-halle.de}
\date{22 January 2003}

\begin{abstract}
A field-theory approach  is used to investigate  the ''spin-glass effects'' on
the critical behaviour of systems with weak temperature-like quenched  disorder.
The  renormalization group (RG) analysis of the effective Hamiltonian of a model
with replica symmetry breaking (RSB) potentials of a general type is carried out
in the two-loop approximation. The fixed-point (FP) stability, recently found
within the one-step RSB RG treatment, is further explored  in terms of replicon
eigenvalues. We find that the traditional FPs, which are usually considered
to  describe the disorder-induced universal critical behaviour, remain stable
when the continuous RSB modes are taken into account.
\end{abstract}


\section{Introduction}
The  spin-glass effects on the critical behaviour of weakly disordered
systems  have received intense attention  during the last several years
[1-7].
It is well-known that the replica method is one of the very few general
analytical methods to investigate disordered systems \cite{parisi_book}.
It enables one to average over disorder and employ standard field theory
techniques, such as the RG analysis. Most of the results obtained using the
RG methods implicitly assume replica symmetry (RS). However, as is well-known from
the large body of works on glass-like systems, the Parisi RSB scheme is
required  to obtain the correct low temperature solutions of  random models
which are characterized by a macroscopic number of ground states
\cite{parisi_book,binder}. Recently, Le Doussal and Giamarchi \cite{le_doussal}
pointed out that although the RG treats fluctuations exactly, it has to be
incorporated with RSB to escape a risk that it will miss the physics
associated with the existence of a large number of metastable states in certain
disordered systems.

Dotsenko {\it et al}  [2-4]  studied the RSB effects on
the critical behaviour of disordered $p$-component magnets to one-loop order
using $\varepsilon$ expansion. They found that for $p<4$ the
FPs, which are usually considered to describe the  disorder-induced
universal critical behaviour, are unstable with respect to the introduction
of RSB potentials. The one-step RSB ansatz \cite{parisi_book} allowed
them to find the new FPs which turned out to be stable within the
one-step RSB  subspace, but the further analysis showed that for a
general type of RSB  there exists no stable FPs and the RG flows
lead to the strong coupling regime \cite{dotsenko_feldman}.
However, the numerous investigations of pure and disordered systems
performed in the two-loop and higher orders of the approximation
prove that the predictions made in the lowest
order of the approximation, especially on the basis of the
$\varepsilon$  expansion, can differ strongly from the real
critical behaviour \cite{our}. Therefore, the results of the RSB
effects investigation in [2-4] must be revised  with the use
of more accurate approximations.

In \cite{prb_2001} the one-step RSB effects on the critical
behaviour have been reconsidered using the field-theory
approach in the two-loop approximation. It was shown that the
RS FPs are stable with
respect to the introduction of the one-step RSB  potentials. However,
it is likely  that the RSB in disordered systems should be of a
general type rather than in the one-step RSB class \cite{wu}.
Therefore, one cannot be sure that FP scenario obtained by means
of the one-step RSB ansatz describes adequately the spin-glass
effects on criticality. In the present work, we re-investigate the
RSB effects on the critical behaviour extending  the field-theoretic
description given in \cite{prb_2001} to the case of the RSB structure
of a general type.

\section{Model and  renormalization}
The starting point of the field-theory approach to the study of spin
systems near critical point in the presence of weak quenched disorder
is the Ginzburg-Landau-Wilson Hamiltonian
\begin{equation}
\fl H=\int \rmd^{d}r \left\{\frac{1}{2}\sum_{i=1}^{p}[\nabla {\phi }_{i}(r)]^{2}+
\frac{1}{2}[\mu_0^2 -\delta \tau (r)]\sum_{i=1}^{p}{\phi }_{i}^{2}(r)+
\frac{1}{4}v \sum_{i,j=1}^{p}{\phi }_{i}^{2}(r){\phi }_{j}^{2}(r) \right\} \label{H}
\end{equation}
where $\phi _{i}(r)$ is the $p$-component order parameter. The quenched
disorder is described by random fluctuations of the effective transition
temperature $\delta \tau(r)$ whose probability distribution is
taken to be symmetric and Gaussian
\begin{equation}
P[\delta \tau]=A \exp\left[-\frac{1}{4u} \int \rmd^{d}r (\delta \tau(r))^2
\right] \label{P}
\end{equation}
where $u$ is the small parameter which is proportional to
the concentration of defects; and $A$ is the normalization constant.
To carry out the averaging over disorder we use  the standard replica
trick. After replicating, we get the effective Hamiltonian
\begin{equation}
\fl H_n=\int \rmd^dr \{\frac{1}{2}\sum_{i=1}^{p}\sum_{a=1}^{n}[\nabla{\phi}%
_{i}^{a}(r)]^{2}+ \frac{1}{2}\mu_0^2\sum_{i=1}^{p}\sum_{a=1}^{n}[{\phi}%
_{i}^{a}(r)]^{2}+ \frac{1}{4}\sum_{i,j=1}^{p}\sum_{a,b=1}^{n}v_{ab}[{\phi}%
_{i}^{a}(r)]^{2}[{\phi}_{j}^{b}(r)]^{2}\} \label{H_n}
\end{equation}
which is a functional of $n$ replications of the original order parameter
with an additional vertex $u$ in the replica symmetric matrix
$v_{ab}=v\delta_{ab}-u$. The properties of the original disordered system
are obtained in the replica number limit $n\rightarrow 0$.

As pointed out in \cite{dotsenko_harris}, near the critical
point the system has a macroscopic number of the local minima
solutions of the saddle-point equation corresponding to the
Hamiltonian (\ref{H}). The traditional RG treatment of the
replicated Hamiltonian (\ref{H_n}) cannot take into account the
existence of these local minima so that the direct application of
the RS RG scheme may be questioned. It was argued that spontaneous
RSB can occur due to the interaction of the fluctuating fields
with the local non-perturbative degrees of freedom coming from the
multiple local minima solutions of the saddle-point equations. It
was shown that the summation over these solutions in the replica
partition function can provide the additional non-trivial RSB
potential $\sum_{a,b}v_{ab}{\phi}_{a}^{2}{\phi}_{b}^{2}$ in which
the matrix $v_{ab}$ has the Parisi RSB structure \cite{parisi_book}.

We now realize the field-theoretic RG analysis of the effective
Hamiltonian (\ref{H_n}) with the RSB potential in the two-loop approximation.
We use the massive field theory scheme with renormalization of the one-particle
irreducible (1PI) vertex functions
$\Gamma_0^{(L,N)}(k_1,..,k_L;q_1,..,q_N;\mu^2_0,\ \{v_{ab}\})$
at non-zero mass and zero external momenta \cite{zinn-justen}.
The 1PI  vertex functions  are defined by
\begin{eqnarray}
\fl \delta(\sum k_i + \sum q_j) \Gamma_0^{(L,N)}(\{k\};\{q\}; \mu^2_0,
\{v_{ab}\}) = \int e^{i(k_i r_i+q_j \hat{r}_j)}  \nonumber\\
\lo \times \langle  \phi(r_1)\dots \phi(r_N) \phi^2(\hat{r}_1)\dots\phi^2(\hat{r}_L)
\rangle_{\rm 1PI} \,\, {\rmd}^d  r_1 \dots {\rmd}^d r_N {\rmd}^d \hat{r}_1 \dots
{\rmd}^d \hat{r}_L \label{vertex}
\end{eqnarray}
where $\{ q\}, \{k \}$ are the sets of external momenta, the brackets $\langle\ldots\rangle$
denote the averaging performed with the effective Hamiltonian (\ref{H_n}),
and for the sake of simplicity the  replica and vectors indexes have been omitted.
The Feynman diagrams that contribute to the vertex functions (\ref{vertex})
involve the momentum integration and depend on the microscopic cut-off $\Lambda_0$.
To study the critical domain which corresponds to the large cut-off limit
($\Lambda_0\rightarrow \infty$) we have to renormalize the theory in order to
absorb the divergences of diagrams in a change of parameters and to obtain meaningful
expressions for the correlation functions. Introducing the renormalization factors for the
fields $Z_{\phi}$, $Z_{\phi^2}$, the renormalized vertex functions $\Gamma_R^{(L,N)}$
are expressed in terms of the bare vertex functions as follows,
\begin{equation}
\Gamma_R^{(L,N)}(\{k\};\{q\};\mu^2, \{g_{ab}\}) = Z_{\phi}^{N/2-L}
Z_{\phi^2}^{L}  \Gamma_0^{(L,N)}(\{k\};\{q\};\mu_0^2,
\{v_{ab}\})  \label{renormalization}
\end{equation}
where $\mu$ is the renormalized mass and $g_{ab}$ is the renormalized matrix
of couplings. To define the renormalization  scheme completely, one has to impose the
renormalization conditions for the renormalized vertex functions:
\begin{equation}
\eqalign{
\left. \Gamma_R^{(0,2)}(k,-k\,;\mu^2,\{g_{ab}\})\right|_{k=0}&=\mu^2   \\
\left. \frac{\partial}{\partial k^2}\Gamma_R^{(0,2)}(k,-k\,;\mu^2,\{g_{ab}\})\right|_{k=0}&=1  \\
\left. \Gamma_{ab\, R}^{(0,4)}(k_1,k_2,k_3,k_4\,;\mu^2,\{g_{ab}\})\right|_{k_i=0}&= \mu^{4-d}g_{ab} \label{conditions1} \\
\left. \Gamma_R^{(1,2)}(k,-k;\,q\,;\mu^2,\{g_{ab}\})\right|_{k=q=0}&=1. }
\end{equation}
The scaling behaviour in the critical domain is described  by the
homogeneous Callan-Symanzik equation for the vertex functions \cite{zinn-justen}:
\begin{eqnarray}
\fl \left[ \mu \frac{\partial}{\partial \mu} +
\sum \limits_{a'b'} \beta_{a'b'}(\{g_{ab}\}) \frac{\partial}{\partial g_{a'b'}} \right.
\nonumber \\
\lo - \left.  \left(\frac{N}{2}-L\right)\gamma_\phi(\{g_{ab}\}){-}
L {\gamma}_{\phi^2}(\{g_{ab}\}) \right]
\Gamma^{(L,N)}_R(\{k\};\{q\};\mu^2,\{g_{ab}\}) = 0 \label{callan}
\end{eqnarray}
with the coefficients given by
\begin{equation}
\fl \beta_{ab}(\{g_{ab}\})=\left. \frac{\partial g_{ab}}{\partial
\ln \mu }\right|_{\{v_{ab}\},\, \mu_0} \,\,
\gamma_{\phi}(\{g_{ab}\})=\left. \frac{\partial Z_{\phi}}{\partial
\ln \mu }\right|_{\{v_{ab}\},\, \mu_0}  \,\,
{\gamma}_{\phi^2}(\{g_{ab}\})=\left. \frac{\partial Z_{\phi^2}}
{\partial \ln \mu }\right|_{\{v_{ab}\},\, \mu_0}. \label{coeff1}
\end{equation}
Equations (\ref{conditions1}) imply that the renormalized matrix $g_{ab}$
has the Parisi RSB structure, which is stipulated by the RSB structure of
the bare matrix $v_{ab}$.
According to the technique of the Parisi RSB
algebra, in the limit $n\rightarrow 0$ the matrix $g_{ab}$ is
parameterized in terms of its diagonal elements $\tilde{g}$ and the
off-diagonal function $g(x)$ defined in the interval $0<x<1$:
$g_{ab}\rightarrow (\tilde{g},g(x))$. The rules for operations
with matrices $g_{ab}$ are  detailed in
\cite{dotsenko_ufn,parisi_book}.  In
\cite{prb_2001}  the renormalization procedure (\ref{renormalization})-(\ref{coeff1})
was carried out with the assumption that all  matrices $g_{ab}$ have
the structure known as the one-step block-like RSB with the function $g(x)$
having the form
\begin{equation} \label{one-step}
  g(x)=\cases{g_{0} & for  $0\leq x<x_{0}$ \\
  g_{1} & for $x_{0}<x\leq 1$}
\end{equation}
where the undetermined step parameter $x_0\in [0,1]$ is related to the
initial disorder distribution. In the present work, we renormalize
the theory  to two-loop order  assuming that the RSB structure is
of the general Parisi type.  The renormalization
conditions (\ref{conditions1}) can be expressed in terms of $\tilde{g}$ and $g(x)$
as follows
\begin{eqnarray}
\left. \Gamma_R^{(0,2)}[k,-k\,;\mu^2,\tilde{g},g(x)]\right|_{k=0}&=&\mu^2 \nonumber \\
\left. \frac{\partial}{\partial k^2}\Gamma_R^{(0,2)}[k,-k\,;\mu^2,\tilde{g},g(x)]\right|_{k=0}&=&1 \nonumber \\
\left. \tilde{\Gamma}_{R}^{(0,4)}[k_1,k_2,k_3,k_4\,;\mu^2,\tilde{g},g(x)]\right|_{k_i=0}&=&
\mu^{4-d}\tilde{g}  \label{conditions2}  \\
\left. \Gamma_{R}^{(0,4)}[k_1,k_2,k_3,k_4;\mu^2,\tilde{g},g(x)]\right|_{k_i=0}&=&
\mu^{4-d}g(x) \nonumber \\
\left. \Gamma_R^{(1,2)}[k,-k\,;q\,;\mu^2,\tilde{g},g(x)]\right|_{k=q=0}&=&1.\nonumber
\end{eqnarray}
Equations (\ref{conditions2}) are simultaneous integral equations, which
show that  we deal with the functional RG. To obtain the renormalized couplings and
the renormalization factors we can solve equations (\ref{conditions2}) perturbatively order by order
in $\tilde{g}$ and $g(x)$.
The functional counterparts of equations (\ref{coeff1}) read
\begin{equation}
\tilde{\beta}[\tilde{g},g(x)]=\left. \frac{\partial \tilde{g}}{\partial
\ln \mu }\right|_{\tilde{v},v(x),\, \mu_0} \,\,
\beta[\tilde{g},g(x)]=\left. \frac{\partial g(x)}{\partial
\ln \mu }\right|_{\tilde{v},v(x),\,\mu_0} \nonumber\\ \label{beta_fun}
\end{equation}
where ($\tilde{v},v(x)$) represents the bare matrix $v_{ab}$. Equations (\ref{beta_fun})
allow us to calculate the $\beta$ functions as functionals of the renormalized parameters
$\tilde{g}$ and $g(x)$. In the two-loop approximation we obtain the following expressions
for the $\beta$ functions,
\numparts
\begin{eqnarray}
\fl \widetilde{\beta }[\tilde{g},g(x)]= \widetilde{\gamma}_1+\widetilde{\gamma}_2+\widetilde{\gamma}_3  \label{beta1} \\
\fl\ \ \ \ \ \ \widetilde{\gamma}_1 = -\tilde{g},
\ \ \widetilde{\gamma}_2 = (8+p){\tilde{g}}^{2}-p{\overline{g^{2}}}  \nonumber \\
\fl\ \ \ \ \ \ \widetilde{\gamma}_3 = ((8\,f-40\,h+20)p+16\,f-176\,h+88){\tilde{g}}^{3}
+ 4\,p\,(6\,h-2\,f-3)\tilde{g}{\overline{g^{2}}} \nonumber \\
\lo + 8\,p(1-2\,h){\overline{g^{3}}}  \nonumber \\
\fl \beta [\tilde{g},g(x)]= \gamma_1+\gamma_2+\gamma_3 \label{beta2}  \\
\fl\ \ \ \ \ \ \gamma_1 = - g(x),
\ \ \gamma_2 = -4g^{2}(x)+(4+2p)\tilde{g}g(x)+2pg(x)+pR_{1}(x)  \nonumber \\
\fl\ \ \ \ \ \ \gamma_3 = 16(1-2h)g^{3}(x)-(8p(1-4h)+ 48 - 96h){\tilde{g}}g^{2}(x)+((8f-48h+28)p \nonumber \\
\lo +16f-48h+24){\tilde{g}}^{2}g(x)+16p(1-2h){\tilde{g}}{\overline{g^{1}}}g(x){+8p(1-2h)\tilde{g}}R_{1}(x)  \nonumber \\
\lo  -8p(1-4h){\overline{g^{1}}}g^{2}(x)+8phg(x)R_{1}(x)+4p(4h-2f-3){\overline{g^{2}}}g(x) \nonumber \\
\lo -8p(1-2h)R_{2}(x) \nonumber
\end{eqnarray}
\endnumparts
where we have introduced $\overline{g^{n}}=\int\limits_{0}^{1}g^{n}(y)\rmd y$,
$R_{1}(x)=\int\limits_{0}^{x}[g(x)-g(y)]^{2}\rmd y$,
$R_{2}(x)=\int\limits_{0}^{x}[g(x)-g(y)][g^{2}(x)-g^{2}(y)]\rmd y$.
The values of two-loop integrals are given by
$f(d=3)=\frac{2}{27}$, $h(d=3)=\frac{2}{3}$, $f(d=2)=0.11464$,
$h(d=2)=0.78129$. Analogous  to [2-5], we changed
$g_{ab}\rightarrow g_{ab}/J$ with $J = \int d^{d}k/(k^2+1)^2$
and $g_{a\neq b}\rightarrow -g_{a\neq b}$ in equations (12). The RS situation
corresponds to the case $g(x) = g_0$ in which equations (12) reduce immediately
to equations obtained previously in \cite{jug} for vertices
$v_1=(p+8)(\tilde{g}+g_0)$ and $v_2=8g_0$.

It is well known that the expansions of RG functions in powers of
coupling constants are asymptotic series. A self-consistent way to
extract the required physical information from the obtained
expressions (12) requires application of special
resummation methods: Borel resummation accompanied by certain
additional procedures (see reference \cite{holovatch} and citation
therein). We employ the Pad\'{e}-Borel resummation method
extended to the functional case. The resummation procedure
consists of several steps: (i) starting from the  functional
$f[\tilde{g},g(x)]$ in the form of  a series  in the auxiliary variable $\theta$,
\begin{equation}
f[\tilde{g},g(x)] = \sum\limits_{m}\gamma_m{\theta}^m \rightarrow \sum\limits_{m}
\frac{\gamma_m}{m!}{\theta}^m \label{borel}
\end{equation}
where $\gamma_{m}$ denotes the contribution of order $g^m$,
we construct its Borel transform (\ref{borel});
(ii) the Borel transform is extrapolated by a rational Pad\'{e}  approximant $[K/L](\theta)$
that is defined as the ratio of two polynomials both in the variable $\theta$
of degrees K and L such that the truncated Taylor expansion of the approximant
is equal to that of the Borel transform of the functional $f$;
(iii) the resummed function $f_{s}$ is then calculated as the inverse Borel transform of
this approximant:
\begin{equation}
f_{s}[\tilde{g},g(x)]=\int\limits_0^\infty d\theta \exp(-\theta)[K/L](\theta).
\end{equation}
In the two-loop approximation $\beta$ functions (12) have
the form $f=\gamma_1\theta+\gamma_2\theta^2+\gamma_3\theta^3$.
Using the [2/1] Pad\'{e} approximant
$[2/1](\theta)= \gamma_1\theta+3\gamma_2^2\theta^2/(6\gamma_2-2\gamma_3\theta)$
we obtain
\begin{equation}
f_{s}[\tilde{g},g(x)]=\gamma_1-\frac{3\gamma_2^2}{2\gamma_3}-\frac{9\gamma_2^3}{2\gamma_3^2}
+ \frac{27\gamma_2^4}{2\gamma_3^3}\exp\left(-\frac{3\gamma_2}{\gamma_3}\right){\rm Ei}
\left(\frac{3\gamma_2}{\gamma_3}\right) \label{beta_s}
\end{equation}%
where $\mathrm{Ei}(x)={\mathcal P.V.}\int_{-\infty }^{x}\rmd t\exp (t)/t$ is the exponential
integral.

\section{Stability of FPs and replicon eigenvalues}
The nature of the critical behaviour is determined by the existence of a
stable FP satisfying the following simultaneous equations
\begin{equation}
\tilde{\beta}_{s}[\tilde{g}^*,g^*(x)]=0, \ \ \  \beta_{s}[\tilde{g}^*,g^*(x)]=0.
\end{equation}
Taking the derivative  over  $x$, we obtain from the last equation
\begin{equation} \label{lambda}
\frac{\rmd}{\rmd x}\beta_{s}[\tilde{g},g(x)] = \lambda[\tilde{g},g(x)]g'(x).
\end{equation}
Equation (\ref{lambda}) implies  that either the FP function $g^*(x)$
has the step-like structure, or it obeys the equation
$\lambda[\tilde{g}^*,g^*(x)]=0$.  The  simplest assumption is that
the one-step RSB (\ref{one-step}) occurs in the weakly disordered
systems \cite{dotsenko_harris} so that the $\beta$ functionals
(12) reduce immediately  to the three functions:
$\beta_{i}(\tilde{g},g_0,g_1)$, $(i=1,2,3)$ obtained in
\cite{prb_2001}. The analysis of the $\beta$ functions shows
that there are three types of non-trivial FPs  for different
values of $p=1,2,3$. Type I with $\tilde{g}^*\neq 0, g_0^* = g_1^*
= 0$ corresponds to the RS FP of a pure system, type II with
$\tilde{g}^*\neq 0, g_0^* = g_1^* \neq 0$ is a disorder-induced RS
FP, and type III with $\tilde{g}^*\neq 0, g_0^* = 0, g_1^* \neq 0$
corresponds to the one-step RSB FP. The parameters  $\tilde{g}^*,
g_0^*,g_1^*$, describing the FPs position, depend on the
coordinate of the step $x_0\in [0,1]$ and  the obtained  values of these
parameters  have been presented changing with  the step $\Delta
x_0 =0.1$ in tables 1-4 of paper \cite{prb_2001}.

The type of critical behaviour of this disordered system for each value
of $p$ is determined by the stability of the corresponding FP.
Within the one-step RSB parameter subspace, the requirement that the FP be
stable reduces to the condition that the eigenvalues $\lambda_i$ of
the matrix
\begin{equation} \label{matrix}
B_{i,j}=\frac{\partial\beta_i(\tilde{g}^*,g_0^*,g_1^*)}{\partial g_j}
\end{equation}
lie in the right complex half-plane. The values $\lambda_i$
calculated in \cite{prb_2001}  show that for three-dimensional (3D) and
two-dimensional (2D) Ising
models ($p=1$) and 3D XY model ($p=2$) the disorder-induced RS
FPs are stable in the one-step RSB subspace, whereas the critical
behaviour of 3D Heisenberg model ($p=3$) is described by the
RS FP of a pure system.

As shown in \cite{dotsenko_feldman} the stability of the FP in the full
RSB space may strongly differ from the stability in the one-step RSB subspace,
so it is quite interesting  to investigate the stability properties of the calculated
early FPs with respect to the RSB modes of a general type. The $\beta$ functions
(12) obtained in the present work give the complete RG description
of system (\ref{H_n}) within the scope of the two-loop approximation.
In order to investigate the full stability properties of FP we have to solve the
RG flow equations
\begin{equation} \label{flow}
\eqalign{
\mu \frac{\partial }{\partial \mu }\tilde{g}_{\mu }=\widetilde{\beta }_{s}[\tilde{g}_{\mu
},g_{\mu }(x)] \nonumber \\
\mu \frac{\partial }{\partial \mu }g_{\mu }(x)=\beta _{s}[\tilde{g}_{\mu },g_{\mu
}(x)]   }
\end{equation}
in the  vicinity of the FP for $\mu\rightarrow 0$ \cite{zinn-justen}.
This task seems to be very complicated in the two-loop
approximation. Recently, an alternative method, the so-called  ''replicon modes'' approach,
was suggested for this complex problem \cite{wu}. This
approach allows for a very simple way to explore the stability
properties with respect to the continuous RSB modes without solving
equations (\ref{flow}) directly. Indeed, the usual way to examine the stability
of the FP of a model with several coupling constants is to linearize  and  diagonalize the RG flow
equations about this FP. The replicon modes approach is the generalization of this method
to the specific system (\ref{H_n}) with the infinite number of coupling constants.
Following works \cite{wu,cesare_1} we introduce the eigenmode function
\begin{equation}\label{Q}
Q(x)=\frac{\rmd}{\rmd x}\left[ g(x)-g^{\ast }(x)\right]
\end{equation}
which characterizes the behaviour of the first derivative of $g(x)$ near the FP in the full
RSB parameter space. In the case of a step-like RSB FP (RS FP, one-step or many-step RSB FP),
as it follows from equation (\ref{lambda}), the flow equation for the eigenmode function has the form
\begin{equation} \label{flowQ}
\mu \frac{\partial }{\partial \mu }Q_{\mu }(x)=\lambda_{\rm rep}(x)Q_{\mu }(x),
\end{equation}%
where  we have introduced the replicon eigenvalues $\lambda_{\rm rep}(x)$ determining
the scaling behaviour of the replicon modes $Q(x)$. Thus, the change from $g(x)$ to $Q(x)$
diagonalizes the flow equation about the step-like RSB FP.
If $\lambda_{\rm rep}(x)$ is positive, $Q_{\mu }(x)$ decreases under the
iteration of the RG transformation and the function $g(x)$ will approach the FP
function $g^*(x)$. If $\lambda_{\rm rep}(x)$ is negative, $g(x)$ will
deviate from its FP expression with a related instability along the continuous
replicon mode directions in the parameter space. If $\lambda_{\rm rep}(x)=0$,
$g(x)$ will keep the initial shape and this should correspond to a marginal
stability.

Using equations (\ref{flow})-(\ref{flowQ}) we derive
$\lambda_{\rm rep}(x)=\lambda[\tilde{g}^*,g^*(x)]$, where
$\lambda[\tilde{g},g(x)]$ is given by equation (\ref{lambda}). Substituting into the latter
the resummed $\beta$ function (\ref{beta_s}) with the coefficients $\gamma_i$ taken from
equation (\ref{beta2}) we obtain the expression for the replicon eigenvalues function
\begin{equation}
\left. \lambda_{\rm rep}(x)=\sum\limits_{m=1}^{3} \left[\frac1{g'(x)}
\frac{\partial \gamma_m}{\partial x}\right]_{\{\tilde{g}^*,g^*(x)\}}
\frac{\partial f_s}
{\partial \gamma_m}\right|_{\{\gamma_i=\gamma_i[\tilde{g}^*,g^*(x)]\}}. \label{replicon}
\end{equation}

The set $\Lambda\equiv\{\lambda_1,\lambda_2,\lambda_3,\lambda_{\rm rep}(x)\}$
composed of the replicon eigenvalues and the eigenvalues of the stability matrix
(\ref{matrix}) yields a complete description of the FP stability properties in
the full RSB parameter space \cite{cesare_1}. We have calculated the replicon
eigenvalues numerically for all FPs, which were  found in the one-step RSB
parameter subspace. Although the coefficient $\gamma_3$ in equation  (\ref{beta2})
depends explicitly on $x$, it turns out that in the case of the one-step RSB FP (\ref{one-step})
the replicon eigenvalues function (\ref{replicon}) depends on $x$ through $g^*(x)$ only, and therefore,
it also has the one-step structure
\begin{equation} \label{one-step_replicon}
  \lambda_{\rm rep}(x)=\cases{
    \lambda_{\rm rep}^{(0)} & for $0\leq x<x_{0}$ \\
    \lambda_{\rm rep}^{(1)} & for $x_{0}<x\leq 1$.}
\end{equation}
We have obtained that the three types of FPs are characterized by the
replicon eigenvalues $\lambda_{\rm rep}^{(0)}$ coinciding with the corresponding
eigenvalues $\lambda_3$ (see tables 1-4 in \cite{prb_2001}) up to the numerical
accuracy. The same identity occurs in
the one-loop approximation \cite{wu,cesare_1}. Nevertheless, there is no
evidence that this identity holds in the higher orders of approximation.
The  obtained values of parameter $\lambda_{\rm rep}^{(1)}$  for the different
types of FPs are presented in table \ref{tab1}.
The analysis of $\Lambda$ (see tables  1-4 of paper \cite{prb_2001} and
table \ref{tab1} of this paper) calculated for the three types of  FPs and for
different values of $p=1, 2, 3$ gives the following results.

{\it (i) 2D and 3D Ising models} ($d=2,3$; $p=1$).  Within the one-step RSB subspace the only stable  FP
is the disorder-induced RS FP (type II)  with  all ${\rm Re}\lambda_i>0$, $i=1,2,3$.
In the case of the 3D Ising model ($d=3$) the disorder-induced RS FP is characterized by the complex
values $\lambda_1$ and $\lambda_2$ leading to RG flows which spiral around the FP
\cite{zinn-justen} and giving rise to the oscillating corrections to scaling \cite{khmelnitskii}.
The type II FP remains stable when the continuous RSB modes are taken into account
($\lambda_{\rm rep}^{(0)},\  \lambda_{\rm rep}^{(1)}>0$). The FP of a pure system
(type I) and the disorder-induced RSB FP (type III) are unstable ($\lambda_3\approx\lambda_{\rm rep}^{(0)}<0$).
Therefore the critical behavior of weakly disordered 2D and 3D Ising systems is realized with
the disorder-induced RS FP.
\Table{Replicon eigenvalues $\lambda_{\rm rep}^{(1)}$.
\label{tab1}}
\br
    &     &  $d=3$    &           &           &   $d=2$   \\ \ns
Type&$x_0$&             \crule{3}             & \crule{1} \\
    &     &  $p=1$    &   $p=2$   &   $p=3$   &   $p=1$   \\ \mr
 I  &     &\-0.169236 &\-0.001673 &  0.131538 &\-0.095180 \\
 II &     &  0.211760 &  0.000003 &  0.014501 &  0.056959 \\
 III& 0.0 &  0.211760 &  0.000003 &  0.014501 &  0.056959 \\
    & 0.1 &  0.245366 &  0.000092 &  0.009909 &  0.066108 \\
    & 0.2 &  0.283705 &  0.000191 &  0.005010 &  0.076207 \\
    & 0.3 &  0.326606 &  0.000302 &\-0.000202 &  0.087391 \\
    & 0.4 &  0.375427 &  0.000425 &\-0.005714 &  0.099813 \\
    & 0.5 &  0.430949 &  0.000566 &\-0.011485 &  0.113656 \\
    & 0.6 &  0.494819 &  0.000728 &\-0.017414 &  0.129128 \\
    & 0.7 &  0.568823 &  0.000913 &\-0.023284 &  0.146478 \\
    & 0.8 &  0.654381 &  0.001130 &\-0.028649 &  0.165992 \\
    & 0.9 &  0.751314 &  0.001388 &\-0.032590 &  0.188004 \\
    & 1.0 &  0.857323 &  0.001695 &\-0.033125 &  0.212903 \\
\br
\endTable

{\it (ii) 3D XY model} ($d=3$; $p=2$).
Although the positive eigenvalues $\lambda_1$,$\lambda_2$,$\lambda_3$,$\lambda_{\rm rep}^{(0)}$,
$\lambda_{\rm rep}^{(1)}$ calculated for the disorder-induced RS FP (type II) indicate that
this FP is stable in the full RSB space, we have reasons to believe that in the higher orders of
approximation the type I FP (which corresponds to the critical behavior of a pure system)
will become stable.  Indeed, the small eigenvalues $\lambda_3,\lambda_{\rm rep}^{(1)}\approx3\cdot10^{-6}$
indicate the weak stability of  the disorder-induced RS FP. On the other hand,
the two-loop RG analysis  of  the disordered systems without RSB potentials \cite{jug}
gives $p_c=2.0114$ for the borderline between regions of stability for the
disorder-induced FP ($p<p_c$) and the FP of a pure system ($p>p_c$), while the  estimation of $p_c$
performed in the six-loop approximation using the pseudo $\varepsilon$ expansion gives
$p_c=1.912$ \cite{holovatch_JPS}. The latter result is in accordance with the Harris criterion,
which suggests that due to the negative value of the specific heat exponent of the pure 3D XY model
the critical behaviour of this model should be stable with respect to the influence of the quenched
disorder at least in the absent of RSB effects.

{\it (iii) 3D Heisenberg model} ($d=3$; $p=3$).
The disorder-induced RS FP is unstable due to the negative  eigenvalues
$\lambda_2,\lambda_3=\lambda_{\rm rep}^{(0)}<0$.
The disorder-induced RSB FP is stable in the directions corresponding to the
continuous RSB modes  for
$0.1\lesssim x_0 \lesssim 0.2$ ($\lambda_{\rm rep}^{(0)}$,$\lambda_{\rm rep}^{(1)}>0$),
but this FP is unstable in the directions corresponding to the one-step RSB modes
for any $x_0$ ($\lambda_{2}<0$).
Additionally, both  FPs are characterized by the non-physical values of coordinate $g_1^*<0$,
so that the RS FP of a pure system is the only stable FP in the full RSB space.
Therefore the influence of weak quenched disorder on the critical behaviour of the 3D Heisenberg model
is irrelevant even with taking into account the RSB effects.

\section{Conclusions}
The RG investigations carried out in the two-loop
approximation show the stability of the critical behaviour of
weakly disordered systems with respect to the introduction of
potentials with the general Parisi RSB structure. In dilute Ising-like
systems, the disorder-induced critical behaviour is realized with RS
FP. The critical behavior of  XY- and  Heisenberg-like systems is not
affected by the weak quenched disorder.

Recently, the phase diagram of the random temperature
Ising ferromagnet was studied within the framework of the Gaussian
variational approximation \cite{dotsenko_tarjus}. The spin-glass
phase separating the paramagnetic and ferromagnetic phases was
found. It was shown that the transition from paramagnetic to
spin-glass state is second order, whereas the transition between
spin-glass and ferromagnetic states is first order. It was also
shown that within the considered approximation there is no RSB in
the spin-glass phase. If it is just the case, we argue that the
paramagnetic--spin-glass transition is controlled by the
traditional RS FP of a weakly disordered system and that slightly
above this critical point the asymptotic behaviour is characterized
by the critical exponents obtained previously for the diluted
Ising model within the RS RG treatment. However, there are some
strong arguments against the existence of an intermediate
spin-glass phase \cite{sherrington} so that to define the nature of
phase transition controlled by the obtained stable RS FP  further
investigation is clearly necessary.

\ack {Support from the
Deutsche Forschungsgemeinschaft (SFB 418) is gratefully 
acknowledged.  I  would also like to thank A. Kudlay for a useful
discussion.}

\end{document}